\documentclass[a4paper,11pt]{article}
\usepackage{pos}
\usepackage{wrapfig}

\title{Search for the Sagittarius tidal stream of axion dark matter around 4.55 $\mu$eV}

\author*[a]{B. R. Ko}
\onbehalf{on behalf of the CAPP-12TB experiment}

\affiliation[a]{Center for Axion and Precision Physics Research, Institute for Basic Science, Daejeon 34051, Republic of Korea}

\emailAdd{brko@ibs.re.kr}

\abstract{We report the first search for the Sagittarius tidal stream
  of axion dark matter around 4.55 $\mu$eV using CAPP-12TB haloscope
  data acquired in March of 2022. Our result excluded the Sagittarius
  tidal stream of Dine-Fischler-Srednicki-Zhitnitskii and
  Kim-Shifman-Vainshtein-Zakharov axion dark matter densities of
  $\rho_a\gtrsim0.184$ and $\gtrsim0.025$~GeV/cm$^{3}$, respectively,
  over a mass range from 4.51 to 4.59~$\mu$eV at a 90\% confidence
  level.  
}

\FullConference{The European Physical Society Conference on High Energy Physics (EPS-HEP2023)\\
 21-25 August 2023\\
Hamburg, Germany\\}

\begin{document}
\maketitle
\section{Introduction}
\begin{wrapfigure}{r}{0.50\textwidth}
\centerline{\includegraphics[width=0.5\textwidth]{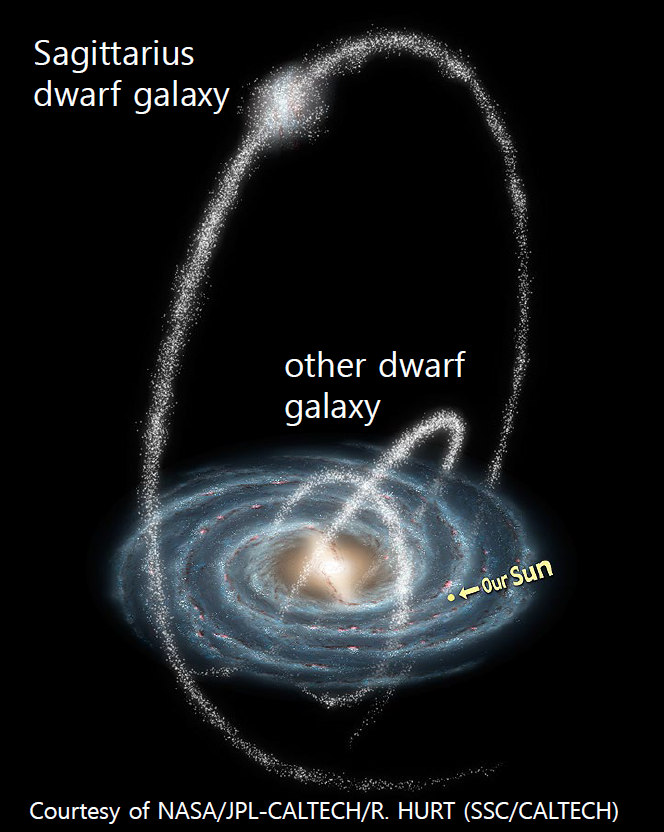}}
  \caption{Milky Way and its satellite galaxies.}
  \label{FIG:DWARF}
\end{wrapfigure}
According to the standard model of Big Bang cosmology and precision
cosmological measurements~\cite{PLANCK}, matter in our Universe makes
up invisible cold dark matter (CDM) approximately of 85\%.
CDM is unknown to date even though there is strong evidence of the
existence of it~\cite{CDM-EVIDENCE} and cannot be handled by the
Standard Model of particle physics (SM).

\begin{figure}[h]
  \centering
  \includegraphics[width=0.48\columnwidth]{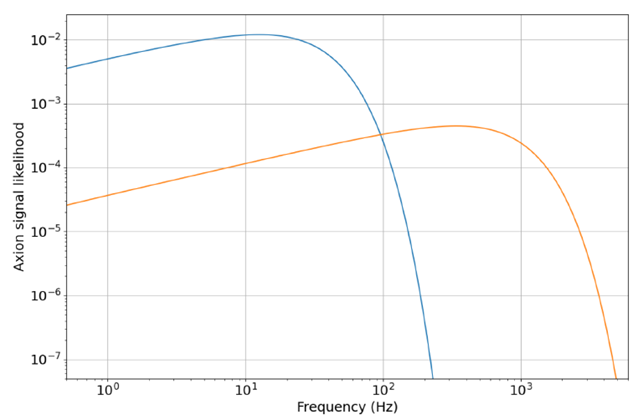} \quad
  \includegraphics[width=0.48\columnwidth]{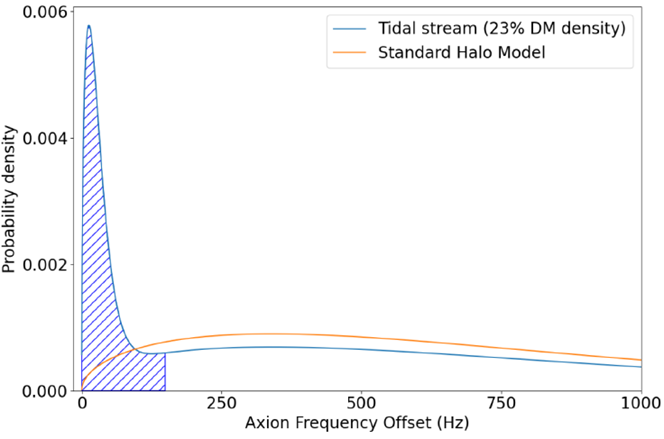} 
  \caption{Axion dark matter signal shapes for a corresponding
    frequency of 1.1 GHz following the SHM (orange lines in the plots)
    and the tidal stream model (blue line in the left plot). The blue
    line in the right plot is the dark matter shape for this work
    assuming axion dark matter constitutes 100\% of the local dark
    matter density, where it contributes 23\% from the tidal stream
    model and 77\% from the SHM. Only the blue hatched region in the
    right plot is considered as the signal region.    
  }
  \label{FIG:DM_MODELS}
\end{figure}

Dark matter in the halo is believed to dominate the local dark matter
density in light of the current galaxy formation and evolution as well
as the strong evidence of the dark matter existence.
However, there could be additional local dark matter contribution from
the Sagittarius dwarf tidal stream according to
K. Freese {\it et al.}~\cite{TIDAL_AXION}, where the Sagittarius dwarf
galaxy is a satellite galaxy of the Milky Way as shown in
Fig.~\ref{FIG:DWARF}.
The tidal force from the Milky Way gravitation pulls the dwarf galaxy
body so that it can form the streams showering onto the Solar System
as also shown in Fig.~\ref{FIG:DWARF}.
Since the Sagittarius dwarf galaxy is also believed to contain its own
dark matter halo, the dark matter from the dwarf galaxy also makes up
the tidal stream and it could contribute at most 23\% of the local
dark matter density~\cite{TIDAL_AXION}. The standard halo model (SHM)
describes the signal shape of dark matter halo in the Milky Way which
is indicated by orange lines in
Fig.~\ref{FIG:DM_MODELS}~\cite{AXION_SHAPE}.
The dark matter signal shape by the tidal stream model is taken from
Ref.~\cite{TIDAL_AXION} and it is the blue line in the left plot of
Fig.~\ref{FIG:DM_MODELS}. Blue line in the right plot of
Fig.~\ref{FIG:DM_MODELS} is the expected dark matter signal shape in
the presence of the tidal stream contribution of 23\% on top of the
dark matter halo contribution of 77\% and is similar to the model in
Ref.~\cite{WIMP_JCAP}, but only the blue hatched region in the right
plot of Fig.~\ref{FIG:DM_MODELS} is taken as the signal region in this
work.
\section{Axion dark matter and axion haloscope}
\begin{wrapfigure}{r}{0.50\textwidth}
  \centerline{\includegraphics[width=0.50\textwidth]{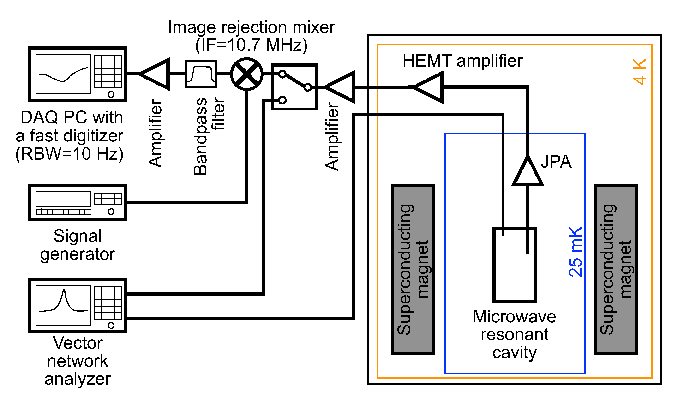}}
  \caption{Outline of an axion haloscope, e.g., the CAPP-12TB haloscope.} 
  \label{FIG:CAPP-12TB}
\end{wrapfigure}
QCD axion or axion~\cite{AXION} was introduced by Peccei and
Quinn~\cite{PQ} to solve the strong CP problem in the
SM~\cite{strongCP} and is predicted to be massive, abundant, and cold
and interacts very weakly with the SM~\cite{CDM_LOW}.
Such axion properties meet the CDM conditions, hence axion is one of
the most appreciated CDM candidates nowadays and referred to as axion
dark matter. The two most popular benchmark models are the
Kim-Shifman-Vainshtein-Zakharov (KSVZ)~\cite{KSVZ} for QCD axions that
couple to beyond the SM heavy quarks, and
Dine-Fischler-Srednicki-Zhitnitskii (DFSZ)~\cite{DFSZ} for those to
the SM quarks and leptons, at tree levels.

A direct axion detection by P. Sikivie~\cite{sikivie} uses the
axion-photon coupling and is known as the axion haloscope. Employing a
high-quality microwave cavity makes the axion haloscope the most
sensitive axion dark matter searches in the microwave region and
Fig.~\ref{FIG:CAPP-12TB} shows a typical outline of the axion
haloscope brought by the CAPP-12TB experiment~\cite{12TB_PRL}.
\section{CAPP-12TB experiment}
The CAPP-12TB experiment depicted in Fig.~\ref{FIG:CAPP-12TB}
constitutes a superconducting solenoid whose central magnetic field is
12 T and bore diameter is 320 mm~\cite{OI}, a 36.85 L frequency
tunable copper cylindrical cavity, a heterodyne receiver chain
with a Josephson Parametric Amplifier (JPA)~\cite{CAPP-JPA} as the
first amplifier, and the fast data acquisition
system~\cite{FDAQ}. With help from a dilution fridge
DRS-1000~\cite{DRS-1000}, the experiment lowered and maintained the
physical temperatures of the cavity and the JPA at around 25 mK.
CAPP-12TB is the second DFSZ axion dark matter sensitive experiment
followed by the ADMX~\cite{ADMX_DFSZ}, where both experiments assumed
the DFSZ axion dark matter halo makes up 100\% of the local dark
matter density, i.e., $\rho^{\rm DFSZ}_a=0.450$ GeV/cm$^{3}$. The
overall operation of the experiment and the relevant measurements can
be found in Ref.~\cite{12TB_PRL}.
\section{Tidal stream dark matter search}
As a complementary dark matter search, this work searched for dark
matter of the tidal stream~\cite{TIDAL_AXION} around 4.55 $\mu$eV
using the same data used for our SHM dark matter
search~\cite{12TB_PRL}, but without a dedicated rescan.
Assuming DFSZ axion dark matter makes up 100\% of the local dark
matter, the aforementioned dark matter model for this work (the blue
hatched region in the right plot of Fig.~\ref{FIG:DM_MODELS}) results
in $\rho^{\rm DFSZ}_a$ of 0.114 GeV/cm$^{3}$.
Table~\ref{TABLE:COMP} summarizes the comparison between our previous
SHM dark matter search~\cite{12TB_PRL} and this tidal stream dark
matter search. Our dark matter model with a signal window of 150 Hz
includes contributions from the tidal stream dark matter of 100\% and
dark matter halo of 2.3\%.
\begin{table}[h]
  \centering
  \begin{tabular}{ | c | c | c | } \hline
    & SHM search & this work \\ \hline
    dark matter  & 100\% DFSZ axion, & 100\% DFSZ axion, \\ 
    constitution & 100\% SHM contribution & 23\% tidal stream and 77\% SHM contributions \\ \hline  
    signal window       & 4050~Hz  & 150~Hz  \\ \hline  
    $\rho^{\rm DFSZ}_a$ & 0.450 GeV/cm$^{3}$ & 0.114 GeV/cm$^{3}$ \\ \hline  
    signal power        & higher ($\because$ higher $\rho^{\rm DFSZ}_a$)& lower ($\because$ lower $\rho^{\rm DFSZ}_a$)\\ \hline
    background   & higher   & lower  \\ 
    fluctuations & ($\because$ wider signal window)  & ($\because$ narrower signal window)  \\ \hline
    cut        & standard (3.718)  & tighter without a rescan  \\ \hline  
  \end{tabular}
  \caption{Comparison between our previous SHM dark matter
    search~\cite{12TB_PRL} and this tidal stream dark matter search.}  
  \label{TABLE:COMP}
\end{table}
\section{Results}
\begin{figure}[h]
  \centering
  \includegraphics[width=0.48\columnwidth]{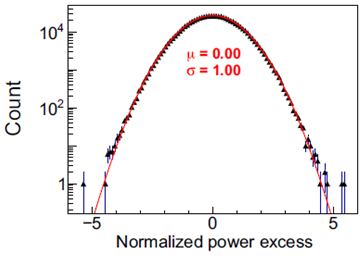} \quad
  \includegraphics[width=0.425\columnwidth]{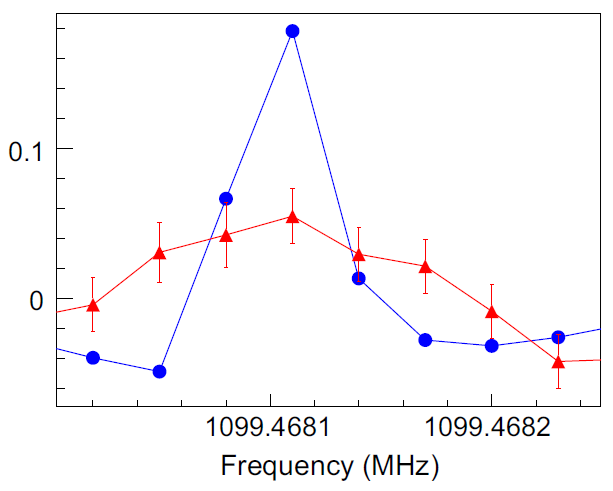} 
  \caption{The left shows the normalized power excess distribution of
    the normalized grand power spectrum and the right shows the test
    statistic (red solid triangles) and alternative hypothesis (blue
    solid circles) that were used for the $\chi^2$ probability test.}  
  \label{FIG:RESULTS}
\end{figure}

The results of this work have been recently published~\cite{12TB_PRD}
and they are summarized in this proceedings.
Data processing followed the usual axion haloscope search analysis
method~\cite{ALL_ANAL}, but applied tighter conditions to exclude all
the excess by a least cut, without a rescan.
A cut of 5.4 was applied to exclude all the excess shown in the left
plot of Fig.~\ref{FIG:RESULTS}. The signal compatibility test was done
for the most significant excess shown in the left plot of
Fig.~\ref{FIG:RESULTS} by the $\chi^2$ probability.
The test statistic and alternative hypothesis are shown in the right
plot of Fig.~\ref{FIG:RESULTS} and the calculated $\chi^2$ probability
of $\mathcal{O}(10^{-9})$ implies that they are not compatible with
each other. The details of the data processing and the associates also
can be found in Ref.~\cite{12TB_PRD}.
\begin{wrapfigure}{r}{0.53\textwidth}
\centerline{\includegraphics[width=0.53\textwidth]{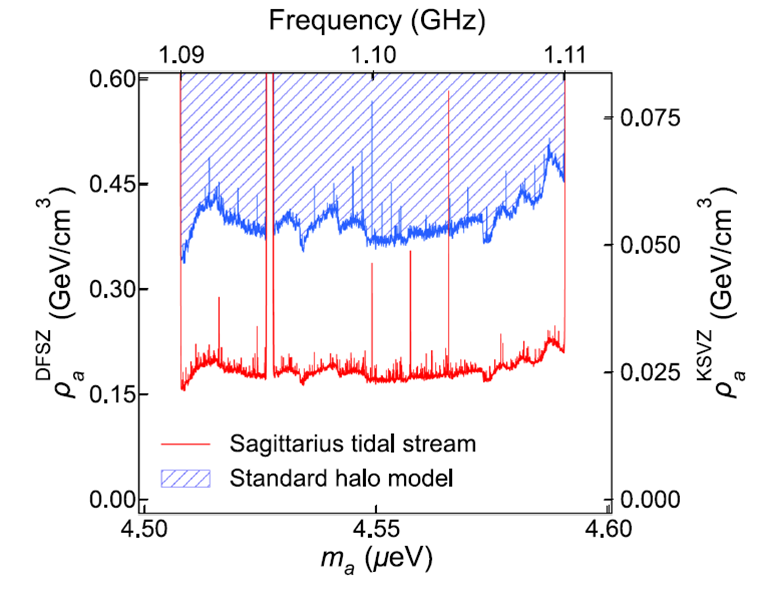}}
  \caption{The blue hatched region shows the exclusion limits for the
    axion dark matter densities from~\cite{12TB_PRL} and the red
    solid line shows those achieved by this work, $\rho^{\rm DFSZ}_a$
    in left and $\rho^{\rm KSVZ}_a$ in right.    
    No results are available around an axion mass of 4.527~$\mu$eV due
    to mode crossing.    
    The spikes are less sensitive frequency points with fewer
    statistics resulting from the filtering
    procedure~\cite{12TB_PRD}.}  
  \label{FIG:LIMITS}
\end{wrapfigure}
In the absence of the axion dark matter signal, Fig.~\ref{FIG:LIMITS}
shows the excluded densities of axion dark matter of the Sagittarius
dwarf tidal stream at a 90\% confidence level (CL).
Dark matter of the tidal streams was ruled out for densities of
$\rho_a\gtrsim0.184$ and $\gtrsim0.025$ GeV/cm$^{3}$ for DFSZ and KSVZ
axions, respectively.
\section{Summary}
CAPP-12TB is the second axion haloscope sensitive to DFSZ axion dark
matter halo. The Sagittarius dwarf tidal stream of axion dark matter
was searched for as a complementary to our previous axion dark matter
halo search~\cite{12TB_PRL}. Although this is a parasitic search
without a dedicated rescan, the result is far beyond KSVZ sensitivity
for axion dark matter from the tidal stream for the first time. This
approach could be extended to the big flow model~\cite{BIG_FLOW} as
done in Ref.~\cite{ADMX_HR} in the near future.
In summary, we excluded the densities of
$\rho^{\rm DFSZ}_a\gtrsim0.184$ GeV/cm$^{3}$ and
$\rho^{\rm KSZV}_a\gtrsim0.025$ GeV/cm$^{3}$ over a mass range from
4.51 to 4.59 $\mu$eV at a 90\% CL.
\acknowledgments
This work was supported by the Institute for Basic Science (IBS) under
Project Code No. IBS-R017-D1-2023-a00, JST ERATO (Grant No. JPMJER
{\bf 1601}), and JSPS KAKENHI (Grant No.~22H04937). A. F. van Loo was
supported by a JSPS postdoctoral fellowship.


\begin{thebibliography}{99}
\bibitem{PLANCK}
  P. A. R. Ade {\it et al.} (Planck Collaboration), Astron. Astrophys. \textbf{594} (2016) A13. 

\bibitem{CDM-EVIDENCE}
  V. Rubin and W. K. Ford Jr., ApJ \textbf{159} (1970) 379;
  Douglas Clowe {\it et al.}, ApJ \textbf{648} (2006) L109.

\bibitem{TIDAL_AXION}
  K. Freese, P. Gondolo, H. J. Newberg, and M. Lewis, Phys. Rev. Lett. \textbf{92} (2004) 111301;
  K. Freese, P. Gondolo, and H. J. Newberg, Phys. Rev. D \textbf{71} (2005) 043516.  
  
\bibitem{AXION_SHAPE}
  Michael S. Turner, Phys. Rev. D \textbf{42} (1990) 3572.  

\bibitem{WIMP_JCAP}
  Chris W. Purcell, Andrew R. Zentner, and Mei-Yu Wang, JCAP \textbf{08} (2012) 027.

\bibitem{AXION}
  S. Weinberg, Phys. Rev. Lett. \textbf{40} (1978) 223;
  F. Wilczek, Phys. Rev. Lett. \textbf{40} (1978) 279.

\bibitem{PQ}
  R. D. Peccei and H. R. Quinn, Phys. Rev. Lett. \textbf{38} (1977) 1440.

\bibitem{strongCP}
  G. 't Hooft, Phys. Rev. Lett, {\bf 37} (1976) 8; 
  Phys. Rev. D {\bf 14} (1976) 3432; {\bf 18} (1978) 2199(E); 
  J. H. Smith, E. M. Purcell, and N. F. Ramsey, Phys. Rev. \textbf{108} (1957) 120;
  W. B. Dress, P. D. Miller, J. M. Pendlebury, P. Perrin, and N. F. Ramsey, Phys. Rev. D {\bf 15} (1977) 9; 
  I. S. Altarev {\it et al.}, Nucl. Phys. \textbf{A341} (1980) 269. 

\bibitem{CDM_LOW}
  J. Preskill, M. B. Wise, and F. Wilczek, Phys. Lett. \textbf{120B} (1983) 127;
  L. F. Abbott and P. Sikivie, Phys. Lett. \textbf{120B} (1983) 133;
  M. Dine and W. Fischler, Phys. Lett. \textbf{120B} (1983) 137.

\bibitem{KSVZ}
  J. E. Kim, Phys. Rev. Lett. \textbf{43} (1979) 103;
  M. A. Shifman, A. I. Vainshtein, and V. I. Zakharov, Nucl. Phys. \textbf{B166} (1980) 493. 

\bibitem{DFSZ}
  A. R. Zhitnitskii, Yad. Fiz. {\bf 31} (1980) 497 [Sov. J. Nucl. Phys. \textbf{31} (1980) 260];
  M. Dine, W. Fischler, and M. Srednicki, Phys. Lett. \textbf{104B} (1981) 199.

\bibitem{sikivie}
  P. Sikivie, Phys. Rev. Lett. \textbf{51} (1983) 1415; Phys. Rev. D {\bf 32} (1985) 2988.

\bibitem{12TB_PRL}
  Andrew K. Yi {\it et al.}, Phys. Rev. Lett. \textbf{130} (2023) 071002.  

\bibitem{OI}
  W. Ma {\it et al.}, IOP Conf. Ser.: Mater. Sci. Eng. \textbf{502} (2019) 012104.  

\bibitem{CAPP-JPA}
  T. Yamamoto {\it et al.}, Appl. Phys. Lett. \textbf{93} (2008) 042510;
  \c{C}a\u{g}lar Kutlu {\it et al.}, Supercond. Sci. Technol. \textbf{34} (2021) 085013.
  
\bibitem{FDAQ}
  S. Ahn {\it et al.}, J. Instrum. \textbf{17} (2022) P05025.

\bibitem{DRS-1000}
  \url{leidencryogenics.nl}.
  
\bibitem{ADMX_DFSZ}
  N. Du {\it et al.} (ADMX Collaboration), Phys. Rev. Lett. \textbf{120} (2018) 151301;
  T. Braine {\it et al.} (ADMX Collaboration), Phys. Rev. Lett. \textbf{124} (2020) 101303;
  C. Bartram {\it et al.} (ADMX Collaboration), Phys. Rev. Lett. \textbf{127} (2021) 261803.

\bibitem{12TB_PRD}
  Andrew K. Yi {\it et al.}, Phys. Rev. D \textbf{108} (2023) L021304.  

\bibitem{ALL_ANAL}
  S. J. Asztalos {\it et al.}, Phys. Rev. D \textbf{64} (2001) 092003;
  B. M. Brubaker, L. Zhong, S. K. Lamoreaux, K. W. Lehnert, and K. A. van Bibber, Phys. Rev D \textbf{96} (2017) 123008;
  S. Ahn, S. Lee,  J. Choi, B. R. Ko, and Y. K. Semertzidis, J. High Energy Phys. \textbf{04} (2021) 297.
  
\bibitem{BIG_FLOW}
  P. Sikivie, Phys. Lett. B \textbf{567} (2003) 1.

\bibitem{ADMX_HR}
  L. Duffy {\it et al.}, Phys. Rev. Lett. \textbf{95} (2005) 091304;
  Phys. Rev. D \textbf{74} (2006) 012006;
  J. Hoskins {\it et al.}, Phys. Rev. D \textbf{84} (2011) 121302(R);
  J. Hoskins {\it et al.}, Phys. Rev. D \textbf{94} (2016) 082001.

\end{thebibliography}
\end{document}